\documentclass[twocolumn,floatfix, showpacs, nofootinbib]{revtex4}

\usepackage[final]{epsfig}
\usepackage{amsmath}
\usepackage{parskip}
 
\def\lsim{\,\raise0.3ex\hbox{$<$\kern-0.75em\raise-1.1ex\hbox{$\sim$}}\,}
\def\gsim{\,\raise0.3ex\hbox{$>$\kern-0.75em\raise-1.1ex\hbox{$\sim$}}\,}
 
\begin{document}
 
\title{Elastic energy loss with respect to the reaction plane in a Monte-Carlo model}
 
\author{Jussi Auvinen}
\email{jussi.a.m.auvinen@jyu.fi}
\author{Kari J.~Eskola}
\email{kari.eskola@phys.jyu.fi}
\author{Hannu Holopainen}
\email{hannu.l.holopainen@jyu.fi}
\author{Thorsten Renk}
\email{trenk@phys.jyu.fi}
\affiliation{Department of Physics, P.O. Box 35, FI-40014 University of Jyv\"askyl\"a, Finland}
\affiliation{Helsinki Institute of Physics, P.O. Box 64, FI-00014, University of Helsinki, Finland}
 

\begin{abstract}
We present a computation of $\pi^0$ nuclear modification factor with respect to the reaction plane in Au+Au collisions at $\sqrt{s_{NN}}=200$ GeV, based on a Monte-Carlo model of elastic energy loss of hard partons traversing the bulk hydrodynamical medium created in ultrarelativistic heavy-ion collisions. We find the incoherent nature of elastic energy loss incompatible with the measured data. 
\end{abstract}

\maketitle

\section{Introduction}

The interaction of hard high transverse momentum ($p_T$) partons with the bulk medium created in heavy-ion collisions, leading to partonic energy loss and jet quenching, is regarded as one of the most important tools to study the properties of the medium. Initially, medium-induced gluon radiation \cite{Jet1,Jet2,Jet3,Jet4,Jet5,Jet6} has been regarded as the most important energy loss mechanism. However, it was pointed out that also elastic energy loss (where the energy from a leading parton is carried away in the recoil of a scattering partner) is a sizeable contribution in perturbative Quantum Chromodynamics (pQCD) \cite{Thoma1,Thoma2,Thoma3,Mustafa,DuttMazumder}. In particular, elastic energy loss seems to help explaining the data on high $P_T$ single electron suppression, widely believed to be driven by heavy-quark energy loss \cite{Djordjevic,Wicks}, for which the radiative energy loss is small due to the dead cone effect \cite{DeadCone}. This raises the question of the importance of elastic processes to light-quark energy loss.

Consistently, in microscopic calculations, a large elastic contribution to the total energy loss was found \cite{Ruppert,MARTINI,Auvinen:2009qm}, under the assumption that the degrees of freedom (DOF) in the medium are perturbatively interacting partons with a thermal spectrum or (almost) free thermal quasiparticles.  In contrast, in a phenomenological top-down analysis of the pathlength dependence of the energy loss seen in the measured data, based on the fact that in a constant medium elastic energy loss increases linearly with pathlength whereas radiative energy loss increases quadratically, a contribution of more than 10\% elastic energy loss to the total was ruled out \cite{ElasticPhenomenology}. These results, however, are not in manifest contradiction, as the assumption of quasi-free scattering partners being available in the medium may not be realized in nature, and as until recently \cite{MARTINI} no comparison with strongly pathlength dependent observables have been performed in microscopical models of elastic energy loss.

The aim of this paper is to resolve this discrepancy by applying a detailed microscopical Monte Carlo (MC) model of elastic energy loss, presented in Ref. \cite{Auvinen:2009qm}, to the nuclear suppression factor as a function of the angle of outgoing hadron relative to the reaction plane, $R_{AA}(\phi)$, for different collision centralities. Our strategy is as follows: We adjust the parameters of our MC energy loss model such that for a given hydrodynamical description, $R_{AA}(\phi)$ data is reproduced for 0-10\% central collisions. Without adjusting further parameters, we then apply the model to non-central collisions. Going from central to peripheral collisions, the mean density of the system drops as well as the mean pathlength a parton travels in the medium. Both effects decrease the amount of suppression (or increase the suppression factor $R_{AA}$), but a linear weighting of the pathlength should lead to less effect than a quadratic weighting. Thus, in an elastic energy loss model, or any incoherent mechanism, the mean $R_{AA}$ increases more slowly from central to peripheral collisions than in a coherent radiative energy loss model. At the same time, for a given centrality the split between emission in the reaction plane and out of the reaction plane is a function of weighted pathlength, so we expect a larger spread for radiative energy loss than for elastic energy loss. Thus, by comparison with the PHENIX data \cite{PHENIX-R_AA-RP,PHENIX-R_AA}, we expect to be able to probe whether the characteristic scaling of elastic energy loss is realized.

The manuscript is organized as follows: First, we describe the hydrodynamical model used for the evolution of the bulk matter. We then give the essential outline of our MC model for elastic parton-medium interactions, followed by a presentation of the results and a discussion.

\section{Hydrodynamics}

We describe the bulk matter as a locally thermalized fluid. For this purpose, we solve the ideal fluid hydrodynamic equations
\begin{equation}
   \partial_\mu T^{\mu\nu} = 0,
\end{equation}
where $T^{\mu\nu} = (\epsilon + P)u^\mu u^\nu - g^{\mu\nu}P$ is the stress-energy tensor and $\epsilon$ is energy density, $P$ is pressure and $u^\mu$ is the fluid flow four-velocity. We also need an Equation of State (EoS) which relates pressure with the energy density and net-baryon number density, $P=P(\epsilon, n_B)$. Our choice here is the EoS from Laine and Schr\"oder \cite{Laine:2006cp}. Since we consider here particle production only at mid-rapidity, we assume the net-baryon density to be negligible. For the same reason, we assume longitudinal boost-invariance. We solve the (2+1)-dimensional numerical problem using the
SHASTA algorithm \cite{Boris,Zalesak}.

As an initial state we use the sWN profile from Ref.~\cite{Kolb:2001qz}. This profile reproduces the centrality dependence of $dN/d\eta$ at mid-rapidity \cite{HolRasEsk} \footnote{When comparing the computed $\frac{dN/d\eta}{N_{part}/2}$ with the data, the same Glauber model must be applied, see Ref. \cite{STAR}.}. Overall normalization is fixed so that we get the same initial entropy in central collisions as in Ref.~\cite{Niemi:2008ta}. Motivated by the EKRT model \cite{Eskola:1999fc} and Ref.~\cite{Niemi:2008ta}, we set the initial time to $\tau_0 = 0.17$~fm. Centrality classes are defined using the optical Glauber model as in Ref. \cite{Niemi:2008ta}.

Thermal spectra for hadrons are calculated using the conventional Cooper-Frye method \cite{Cooper}, where particle emission is calculated from a constant-temperature surface. The freeze-out temperature $T_{\text{dec}}=160$~MeV is fixed so that we reproduce the measured $p_T$ spectrum of pions \cite{Adler:2003cb} up to $p_T \approx 2$~GeV at various centralities \cite{HolRasEsk}. Strong and electromagnetic two- and three-particle 
decays are taken into account before comparing with experimental data.

\begin{figure}[t]
   \includegraphics[height=9.0cm]{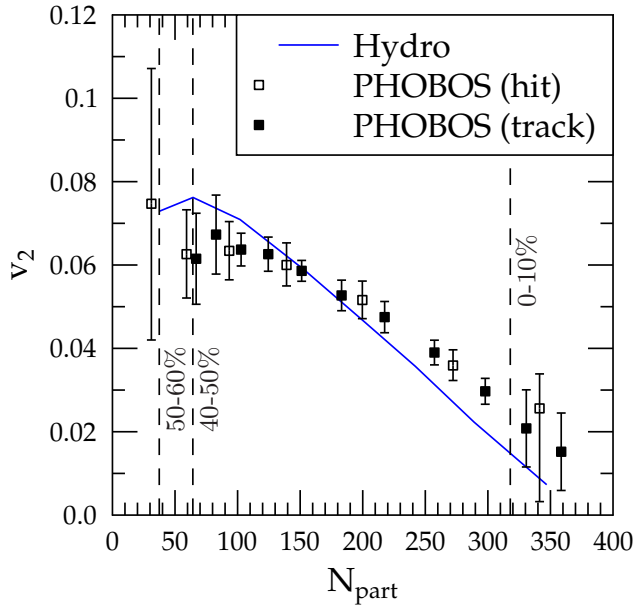}
   \caption{\protect\small (Color online) Integrated elliptic flow for Au+Au collisions at $\sqrt{s_{NN}} = 200$~GeV. Data are from the PHOBOS collaboration \cite{Alver:2006wh}. The data are shown with statistical and systematic errors added in quadrature.}
   \label{fig: v2 int}
\end{figure}

Elliptic flow, which measures the momentum anisotropy of final state particles, is produced during hydrodynamical evolution since the initial state has spatial anisotropy. Because of this anisotropy, also the average pathlength of hard partons is angle dependent. So, to make reasonable energy loss calculations, it is important to constrain the initial eccentricity from the bulk matter elliptic flow.
In Fig.~\ref{fig: v2 int} we show the integrated elliptic flow. We do not get enough elliptic flow in central collisions, but, as explained in \cite{Holopainen:2010gz}, this is due to the lack of initial state density fluctuations and a different reference plane definition than in the experimental analysis. These results show that we can reproduce the bulk observables reasonably well, so that we can do meaningful energy loss calculations using this hydrodynamical model as background.

\section{The Monte Carlo simulation}

We model the energy loss of a hard parton by incoherent partonic $2\rightarrow$ 2 processes in pQCD, with scattering partners sampled from the medium. Our simulation of energy losses of high-energy partons in the produced QCD matter is based on the scattering rate for a high-energy parton of a type $i$, 
\begin{equation}
\label{totgamma}
\Gamma_i (p_1,u(x),T(x)) = \sum_{j(kl)} \, \Gamma_{ij\rightarrow kl}(p_1,u(x),T(x)),
\end{equation}
where we account for all possible partonic processes $ij\rightarrow kl$ by summing over all types of collision partners, $j=u,d,s,\bar u, \bar d, \bar s,g$ in the initial state, and over all possible parton type pairs $(kl)$ in the final state. In general, the scattering rate depends on the frame, and in particular on the high-energy parton's 4-momentum $p_1$, on the flow 4-velocity $u(x)$ and on the temperature $T(x)$ of the fluid at each space-time location $x$. 

In the local rest-frame of the fluid, we can express the scattering rate as follows \cite{Auvinen:2009qm}:

\begin{equation}
\label{scattrate}
\Gamma_{ij\rightarrow kl} = \frac{1}{16\pi^2E_1^2}\int_{\frac{m^2}{2E_1}}^{\infty}dE_2f_j(E_2,T) \Omega_{ij\rightarrow kl}(E_1,E_2,m^2),
\end{equation}
where
\begin{equation}
\label{omegafunction}
\Omega_{ij\rightarrow kl}(E_1,E_2,m^2)=\int_{2m^2}^{4E_1E_2}ds [s\sigma_{ij\rightarrow kl}(s)].
\end{equation}

Here $E_1$ is the energy of the high-energy parton $i$ in this frame and $E_2$ is the energy of the thermal particle $j$ with a distribution function $f_j(E_2,T)$, which is the Bose-Einstein distribution $f_g = f_B(E_2,T)$ for gluons and the Fermi-Dirac distribution $f_q = f_D(E_2,T)$ for quarks. 
The scattering cross section $\sigma_{ij\rightarrow kl}(s)$ appearing in \eqref{omegafunction} depends on the standard Mandelstam variable $s$. A thermal-mass-like overall cut-off scale $m=s_mg_sT$ is introduced in order to regularize the singularities appearing in the cross section when the momentum exchange between partons approaches zero. Here $g_s$ is the strong coupling constant and $s_m$ is a parameter of the order of one. Since we cannot meaningfully control the running of $\alpha_s = \frac{g_s^2}{4\pi}$, we keep the strong coupling constant fixed with momentum scale.

To initiate the hard massless parton of a type $i$ in each event, we sample the LO pQCD single-jet production spectrum (for more details, see \cite{Eskola:2002kv}) at $y_i=0$ and at $p_{Tmin}\leq p_T\leq \sqrt s/2$. For the parton distribution functions (PDFs), we use the CTEQ6L1 set \cite{CTEQ}. For now, we neglect the nuclear effects to the PDFs \cite{NPDF,EKS98,EPS09}, since these are small in comparison with the ones arising from the final state interactions with the medium. The initial rapidity \(y_i\) is randomly generated in the range \([y_{min},y_{max}]\) from a flat distribution. This fixes the hard-parton energy \(E\) and polar angle \(\theta\) of its momentum vector ${\bf p}=(p_x,p_y,p_z)$. The azimuth angle \(\phi\), defined with the reaction plane, is evenly distributed between \([0,2 \pi]\). 

The hard parton is assumed to start interacting with the medium at the initial longitudinal proper time \(\tau_0\) of our hydrodynamical model. Since in the c.m. frame all hard partons are produced in the Lorentz-contracted overlap region at $z\approx 0$, the longitudinal position at later times (before the first collision at $\tau\ge \tau_0$) is assumed to be determined by the longitudinal momentum only. The initial time and longitudinal coordinates for the hard parton are thus \(t_0 = \tau_0 \cosh{y_i}\) and \(z_0 = \tau_0 \sinh{y_i} \). The coordinates on the transverse plane in the beginning of the simulation are then $x_0=x_i+\frac{p_x}{E}t_0$ and $y_0=y_i+\frac{p_y}{E}t_0$, where the parton position in the transverse plane at $t=0$, $(x_i,y_i)$, is sampled from the nuclear overlap function
\begin{equation}
\label{taa}
T_{AA}({\bf b})= \int d^2 {\bf s} \, T_A({\bf s}+ {\bf b}/2) T_A({\bf s}-{\bf b}/2),
\end{equation}  
where \({\bf b}\) is the impact parameter. The nuclear thickness function \(T_A({\bf s})\) is defined as usual, 
\begin{equation}
\label{nucthick}
T_A({\bf s}) = \int dz \, \rho_A({\bf s},z),
\end{equation}
where for the nuclear density \(\rho_A({\bf r})\) we use the standard Woods-Saxon distribution for the gold nucleus. 

The hard parton propagates through the plasma in small time steps \(\Delta t\), during which we propagate the parton in position space. The probability for not colliding in this time interval is assumed to be given by the Poisson distribution
\begin{equation}
P(\text{No collisions in } \Delta t)=e^{-\Gamma_i \Delta t}, 
\end{equation}
where \(\Gamma_i\) is the total scattering rate \eqref{totgamma} for the hard parton of the type $i$. Hence the probability to collide at least once during the time \(\Delta t\) is \(1-e^{-\Gamma_i \Delta t} \approx \Gamma_i \Delta t+ {\cal O}((\Gamma_i\Delta t)^2)\). For small enough \(\Delta t\) we can assume that there will be at most one collision. Note that since we calculate the scattering rates \eqref{totgamma} in the local rest frame of the quark-gluon plasma fluid element, we must boost also the time step \(\Delta t\) to the same frame. Should a scattering happen, the probability $P_{ij\rightarrow kl}$ for a given type of scattering process is determined by the ratios of the partial scattering rates \eqref{scattrate} to the total scattering rate \eqref{totgamma}. After scattering, the final state parton with highest energy is chosen as the new hard parton to be propagated further, for which we repeat the procedure outlined above with the next timestep. 

We take into account the system's slow transformation from quark-gluon plasma to hadron gas by using an effective temperature
\begin{equation}
\label{efftemp}
T_{eff}=\left(\frac{30}{g_Q \pi^2}\epsilon\right)^{1/4},
\end{equation}
where \(g_Q=g_g + \frac{7}{8}2N_fg_q = \frac{95}{2}\) is the quark-gluon plasma degrees of freedom with gluon and quark DOF being $g_g=16$ and $g_q=6$, respectively, and number of quark flavors $N_f=3$. Our simulation ends at \(T=165\) GeV, when the system can be considered to have completely transformed into hadron gas. We assume no significant interaction between the high-energy parton and the fully hadronic medium.

The outcome of the procedure described above is a medium-modified distribution of high-energy partons, $\frac{dN^{AA \rightarrow f+X}}{dp_T dy}$. In order to calculate the nuclear modification factor for neutral pions,

\begin{equation}
R_{AA}(P_T,y,{\bf b}) = \frac{dN^{\pi^0}_{AA}/dP_Tdy}{T_{AA}({\bf b}) d\sigma^{pp}/dP_Tdy},
\end{equation}

we hadronize this distribution by convoluting the obtained partonic distribution with the fragmentation function \(D_{f \rightarrow \pi}(z,\mu_F^2)\):
\begin{equation}
\label{hadrondistr}
\begin{split}
\frac{dN^{AA \rightarrow {\pi^0}+X}}{dP_T dy} = &\sum_f \int dp_T dy \frac{dN^{AA \rightarrow f+X}}{dp_T dy}\cdot\\
&  \int_0^1 dz D_{f \rightarrow \pi} (z,\mu_F^2) \delta(P_T-zp_T),
\end{split}
\end{equation}
where \(z = \frac{P_T}{p_T}\) is the fraction of the final parton momentum \(p_T\) available to the hadron with momentum $P_T$, and \(\mu_F \sim P_T\) is the fragmentation scale. The Kniehl-Kramer-P\"{o}tter (KKP) fragmentation functions \cite{KKP} have been used for this work. 

\section{Results}

\begin{figure*}[t]
\centering
\includegraphics[width=18cm]{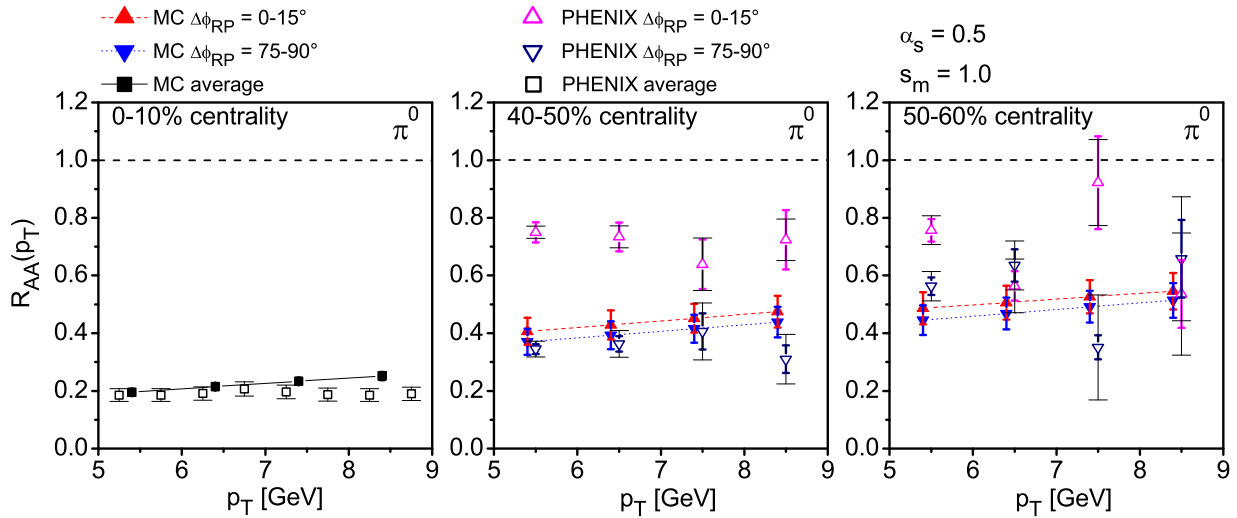}
\caption{(Color online) Left panel: The \(\pi^0\) nuclear modification factor, averaged over the reaction plane angle, for 0-10\% centrality. Middle and right panel: The \(\pi^0\) nuclear modification factor dependence on the reaction plane angle $\Delta \phi$ for 40-50\% (middle panel) and 50-60\% centrality (right panel). Only the smallest- and the largest-angle bins are shown. The strong coupling constant is \(\alpha_s=0.5\) and mass parameter $s_m=1.0$. The simulation points (solid squares and triangles) are connected with lines to guide the eye. PHENIX data is from \cite{PHENIX-R_AA} (0-10\% centrality, open squares) and \cite{PHENIX-R_AA-RP} (40-50\% and 50-60\% centrality, open triangles). Colored bars with small cap represent statistical errors; black bars with wide cap are systematic errors.}
\label{graph_raaphi3}
\end{figure*}

In the following, our interest is in the neutral pions with high transverse momentum $P_{T_{min}} \geq 5.0$ GeV, to account for the observation that at low $P_T$ the main hadron production mechanism is not independent fragmentation. We average all observables across the rapidity window $[-0.35,0.35]$ which corresponds to the PHENIX acceptance. Due to the non-eikonal propagation of the hard partons (see discussion in Ref. \cite{Auvinen:2009qm}) in our simulation, the initial rapidity window is $y_{min}=-y_{max}=-1$ in order to account for all the possible partons falling into the final rapidity window. As an attempt to emulate the reaction plane measurements by PHENIX \cite{PHENIX-R_AA-RP}, the high-momentum partons within the final rapidity acceptance have been divided in six bins of angle $\Delta \phi$ with respect to the reaction plane, from $\Delta \phi = 0-15^\circ$ to $\Delta \phi = 75-90^\circ$.

For this work, the value of strong coupling constant is chosen to be $\alpha_s = 0.5$, which produces roughly the right amount of suppression to match the measured nuclear modification in the 0-10\% centrality bin. In effect, such tuning emulates also the incoherent higher-order processes, whose contribution can be significant. The idea is then to examine how well our tuned MC model agrees with the measurement in other centralities. As our interest here is particularly in the dependence on the reaction plane angle, we concentrate on the more peripheral cases where the effect should be visible best. The simulation results for 0-10\%, 40-50\% and 50-60\% centrality bins, compared with the measurements, are shown in Fig. \ref{graph_raaphi3}. 

It is clear from the figure that our model cannot reproduce the reaction plane angle dependence seen in the PHENIX experiment. Also the inclusive, angle-averaged nuclear modification factor fails to match with the experimental data: The computed suppression decreases too slowly as one advances to the more peripheral collisions.

\section{Discussion}

Our result, obtained for $R_{AA}(\phi)$ in a microscopical well-controlled MC model with a realistic hydro background, is an important one since it demonstrates that a purely incoherent energy-loss framework seems to be in a striking  contradiction with the present RHIC data. In essence, we confirm the findings of the phenomenological analysis \cite{ElasticPhenomenology} that the pathlength dependence of elastic energy loss is incompatible with the data. Taken together with evidence that even the quadratic pathlength dependence of coherent radiative energy loss in pQCD may face some challenge in describing the data \cite{Jia}, this implies that any contribution of elastic energy loss must be rather small (we refrain from trying a combined fit of coherent radiative and incoherent elastic energy loss to the data at this point, but 10\%, as obtained in Ref. \cite{ElasticPhenomenology}, seems certainly reasonable).

The implication is that the main assumption made in pQCD calculations of elastic energy loss, i.e. that the medium DOF are almost free (quasi)-particles which can take a sizeable amount of recoil energy away from a leading parton may not be true in nature. The assumption of static color dipoles (leading to purely radiative energy loss), difficult as it may be to justify, seems to work much better in explaining the data. Given that also a cubic pathlength dependence is consistent with the data \cite{AdS}, it is unfortunately too early to draw a firm conclusion with regard to the DOF and the energy loss mechanism realized in the medium; more systematic studies need to be done. However, given that an ideal fluid description of the bulk medium, which has a zero mean free path assumption implicitly made, works well to describe the bulk, it should perhaps not come as a surprise that an infinite mean free path assumption (the ideal parton gas approximation for $f_j$ in Eq.\ \eqref{scattrate}) does not work well to describe the same medium as seen by a hard parton.

At this point, it is also not clear what the implications of our findings for the question of heavy-quark energy loss are. The reason is that heavy quarks are rare probes, i.e. they are not thermally excited in the medium. While a reaction like $qg \rightarrow qg$ in which the incoming quark is hard, but the outgoing gluon carries 90\% of its energy must be counted as 10\% energy loss from the leading parton if the incoming quark is light, a $c$ quark is always uniquely identified by its flavour and the same reaction must be interpreted as 90\% energy loss from the tagged heavy quark. Thus, elastic energy loss is generically more efficient for heavy quarks than for light quarks, and we will investigate the question of heavy quark elastic energy loss in a future work.

In this work the initial state density fluctuations were neglected. Because of this, all the results were determined using the reaction plane, which is defined by the impact parameter. However, the experimental results for $R_{AA}(\phi)$ are calculated with respect 
to event plane, which may differ from the reaction plane. At least at low $p_T$ this difference is important when calculating $v_2$ \cite{Holopainen:2010gz}. To study whether the initial state density fluctuations and a reference plane definition consistent with the experiment are as important in 
$R_{AA}(\phi)$ is left for future work.

\begin{acknowledgments}
J.A. gratefully acknowledges the grant from the Jenny and Antti Wihuri Foundation and H.H. the financial support from the national Graduate School of Particle and Nuclear Physics. T.R. is supported by Academy Research Fellowship of the Academy of Finland, Project 130472. Additional support comes from K.J.E.'s Academy Project 133005.
\end{acknowledgments}

\end{document}